# Spin-valve Effect in NiFe/$MoS_2$/NiFe Junctions

*Weiyi Wang*[1,2], *Awadhesh Narayan*[3,4], *Lei Tang*[1], *Kapildeb Dolui*[5], *Yanwen Liu*[1,2], *Xiang Yuan*[1,2], *Yibo Jin*[1], *Yizheng Wu*[1,2], *Ivan Rungger*[3], *Stefano Sanvito*[3], *Faxian Xiu*[1,2,†]

[1]State Key Laboratory of Surface Physics and Department of Physics, Fudan University, Shanghai 200433, China

[2]Collaborative Innovation Center of Advanced Microstructures, Fudan University, Shanghai 200433, China

[3]School of Physics, AMBER and CRANN Institute, Trinity College, Dublin 2, Ireland

[4]Department of Physics, University of Illinois at Urbana-Champaign, Urbana, IL 61801, USA.

[5]Graphene Research Center and Department of Physics, National University of Singapore, 2 Science Drive 3, 117542, Singapore

[†]Correspondence and requests for materials should be addressed to F.X. (E-mail: faxian@fudan.edu.cn)

## Abstract

**Two-dimensional (2D) layered transition metal dichalcogenides (TMDs) have been recently proposed as appealing candidate materials for spintronic applications owing to their distinctive atomic crystal structure and exotic physical properties arising from the large bonding anisotropy. Here we introduce the first $MoS_2$-based spin-valves that employ monolayer $MoS_2$ as the nonmagnetic spacer. In contrast with what expected from the semiconducting band-structure of $MoS_2$, the vertically sandwiched-$MoS_2$ layers exhibit metallic behavior. This originates from their strong hybridization with the Ni and Fe atoms of the Permalloy (Py) electrode. The spin-valve effect is observed up to 240 K, with the highest magnetoresistance (MR) up to 0.73% at low temperatures. The experimental work is accompanied by the first principle electron transport calculations, which reveal an MR of ~ 9% for an ideal Py/$MoS_2$/Py junction. Our results clearly identify TMDs as a promising spacer compound in magnetic tunnel junctions and may open a new avenue for the TMDs-based spintronic applications.**

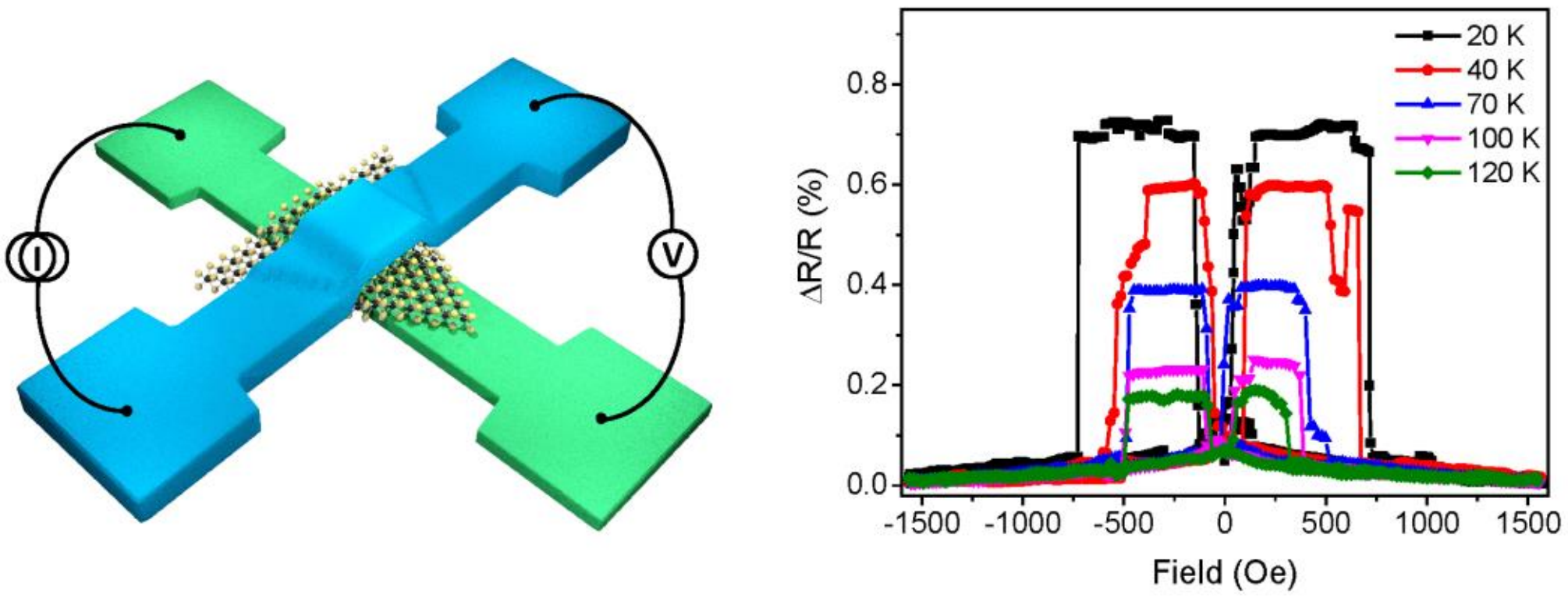

Spin-valve devices have been the subject of intense study over the last few years as they hold the promise of realizing the next-generation of magnetic random-access-memories (MRAM)[1–3], magnetic sensors[4,5] and novel programmable logic devices[6,7]. In the most common situation the non-magnetic spacer in a spin-valve is a wide-gap oxide, usually $Al_2O_3$[8–10] or MgO[11–15]. Recently, 2D materials, with atomically-thin layers, have been proposed as useful nonmagnetic spacer layers in spin-valve devices.[16–18] Soon after such predictions, graphene[19–21] and Boron Nitride[22] (BN) were exploited as the tunnel layers in magnetic tunnel junctions (MTJ) producing considerable tunnel magnetoresistance (TMR) signals with good stability.

In contrast to graphene and BN, which are respectively a metal and a wide-gap semiconductor, the emerging transition metal dichalcogenides (TMDs) family displays a large variety of electronic properties ranging from semi-conductivity to half-metallic magnetism, to superconductivity[23]. As exemplary representative of the TMDs, $MoS_2$ has a tunable bandgap that changes from indirect, 1.2 eV, in the bulk to direct, 1.8 eV, for monolayers, yielding conventional semiconducting characteristics[24–30]. However, there are only a few experimental studies on the vertical transport properties of $MoS_2$ such as the resonant tunneling[31]. As the adjacent layers in 2D material crystals are weakly bound through van der Waals interactions, the weak interlayer coupling and wave function overlap produce relatively poor conductivity in the vertical direction[23]. A giant spin splitting induced by spin-orbit coupling was also found in monolayer $MoS_2$ due to the loss of the inversion symmetry of the bulk compounds, which tends to stabilize the spin polarization in the out-of-plane direction[32,33]. Importantly, theory predicts that $MoS_2$ can produce reasonable magnetoresistance when sandwiched between two Fe electrodes owing to the efficient spin injection[18], thus rendering the possibility of realizing tunable spin-valve devices based on 2D-TMDs. Despite the extensive theoretical efforts, such proof-of-concept devices remain unexplored experimentally.

In this paper, we report on the first fabrication and characterizations of $MoS_2$-based spin-valve junctions. Atomically-thin $MoS_2$ is sandwiched between two Permalloy (Py, $Ni_{80}Fe_{20}$) electrodes and serves as the nonmagnetic spacer layer. The junctions show spin-valve signals up to 240 K with an MR of ~0.4% at 10 K. Upon the reduction of the surface oxidation of the Py by capping a thin gold layer, the MR is dramatically improved to 0.73%. These experimental results are supported by our *ab initio* electron transport calculations, which predicts a magnetoresistance value of ~ 9% for an ideal junction setup.

High-quality $MoS_2$ monolayers were grown on a $SiO_2$ (270 nm)/Si substrate by chemical vapor deposition (CVD) (see inset of Figure 1a). Both photoluminescence (PL) and Raman spectra were taken to identify the thickness. The peak in the PL spectrum is positioned near 1.87 eV (Figure 1a). The Raman shift between two vibration modes, *i.e.*, the in-plane mode $E^1_{2g}$ and the out-of-plane mode $A_{1g}$, is about 19.8 $cm^{-1}$ (Figure 1b). Both experiments thus suggest that our $MoS_2$ is indeed a monolayer[34,35]. A schematic diagram of the devices is shown in Figure 1c. The devices were measured with a four-terminal setup, where the current flows perpendicular to the device plane (CPP geometry) of the junction[36]. The magnetic field is applied in-plane along the bottom electrode as shown in Figure 1d. The junction area is estimated to be 2 μm$^2$, and the resistance-area (RA) products of all the samples are on the order of $10^{-10}$ Ω·m$^2$.

The spin-valve effect is revealed by the magnetoresistance, namely the junction resistance depends on the orientation of the macroscopic magnetization of the ferromagnetic (FM) electrodes. If the magnetization vectors of the two FM layers are parallel to each other, the device is in a low resistance state, $R_P$, otherwise in a high resistance one, $R_{AP}$. The associated magnetoresistance is defined as $MR = (R_{AP} - R_P)/R_P$, which measures the relative resistance change or the effect amplitude. Generally, there are two possible ways to obtain an antiparallel alignment. The first is to use ferromagnets with different coercive fields, either by using different materials or different film thicknesses. The other is to pin one of the ferromagnets to an antiferromagnet. The former one is preferred here because of the simpler fabrication process involved. In order to generate a large difference in coercivity between the two FM electrodes, the thicknesses of the bottom and top Py were designed to be 20 and 50 nm, respectively. A thin Co layer of 10 nm was also deposited beneath the bottom Py electrodes to further increase the difference in coercivity. Thus the bottom electrode can be regarded as the "fixed layer" while the top one is termed "free layer".

Figure 2a displays the resistance of one of our Py/$MoS_2$/Py/Co junctions as a function of temperature ($R$-$T$). We note that the resistance decreases with reducing temperature, unveiling an unexpected metallic behavior of the junction despite the semiconducting nature of $MoS_2$. Our density functional theory calculations elucidate that this is the consequence of the strong hybridization between the interface S atoms and the Fe/Ni atoms of the Py, which leads to the junction exhibiting a metallic transport behavior (see Figure 4 and the theoretical section in supplementary information). The perfect linear current-voltage (*I*-*V*) curves in Figure 2a inset reveal an Ohmic contact between $MoS_2$ and the Py electrodes. Both the metallic *R*-*T* behavior and the linear *I*-*V* characteristics suggest that the monolayer $MoS_2$ behaves as a conducting thin film rather than a tunnel barrier between the two FM electrodes.

The temperature-dependent MR for a representative Py/$MoS_2$/Py/Co junction is shown in Figure 2b. As the magnetic field scans from -2000 to 2000 Oe, the alignment of the top and bottom electrodes switches from parallel to antiparallel, and finally back to parallel, resulting in the observation of a resistance plateau. Such spin-valve signals are persistent up to 240 K. The MR of the junction is determined to be 0.4% at 10 K.

We may analyze the obtained MR by using a simple injection and diffusion model[37]. In conventional MTJ devices, assuming the conservation of the spins in the tunneling process, the Julliere's model[38] has been often used to describe the tunnel magnetoresistance with

$$TMR = \frac{2P_1P_2}{1-P_1P_2}, \tag{1}$$

where $P_1$ and $P_2$ are the electron spin polarizations of the two FM metals. In the case in which the interlayer is conducting, *i.e.* the process is not tunneling-like, the current polarization, $P_1$, decays as $\exp(-z/\lambda_S)$,[37] where $z$ is the diffusion distance along the normal direction and $\lambda_S$ is the spin diffusion length in the nonmagnetic interlayer. Thus the MR can be approximated as

$$MR = \frac{2P_1P_2e^{-d/\lambda_S}}{1-P_1P_2e^{-d/\lambda_S}}, \quad (2)$$

where $d$ is the length of the characteristic diffusive trajectories of the electrons, which could be roughly estimated to be on the same order as the thickness of the monolayer $MoS_2$. As $d$ is far less than $\lambda_S$, the surviving probability, $\exp(-d/\lambda_S) \approx 1$. Assuming that $P_1 \approx P_2 = P$ as suggested by the fact that the two electrodes have the same nominal composition, then the polarization, $P$ , of the two Py electrodes is estimated to be ~ 4.5%. Figure 2c shows the MR amplitude versus temperature. As expected, the MR ratio decreases from 0.4% at 10 K to 0.2% at 240 K. The MR reduction at elevated temperature can be attributed to many possible reasons such as impurity scattering and thermal smearing of the electron energy distribution[39]. Similar spin-valve signals have also been obtained in other devices with identical device structure, evidencing a good reproducibility (see supplementary Figure S2).

Since the bottom Py electrodes are exposed to air during the fabrication process, surface oxidation of the Py is inevitable. In order to circumvent this issue, we have re-designed the junctions stack by looking at the following architecture Py/$MoS_2$/Au/Py/Co. Here, a very thin capping layer of Au (~ 2 nm) was *in-situ* sputtered over the Py to prevent surface oxidation. The *R-T* curve of the junction also exhibits a metallic behavior (Figure 3a), and the *I-V* curves are still linear (Figure 3a inset). This suggests that the Au layer does not change much the overall resistance. As expected, the MR increases to 0.73% owing to the absence of an oxide layer on the Py electrodes. The associated spin polarization, $P$, of the FM/$MoS_2$ surface is also improved to 6.2%. In order to rule out the possibility of being the Au capping layer to determine the MR, we have fabricated a control device with Py/Au/Py/Co structure, *i.e.* without sandwiching $MoS_2$ between the top and bottom electrodes. As shown in Figure S1c of the supplementary information, the control sample does not exhibit apparent spin valve signal, pointing to the fact that the spin filtering effect originates exclusively from the $MoS_2$ monolayer. Finally, the magnitude of the MR decreases from 0.73% at 20 K to 0.2% at 120 K (Figure 3b), similarly to what found for the sample that does not include the Au capping layer.

The decrease of the MR magnitude with temperature (Figure 3c) is analyzed by fitting the data to the Bloch's law, where the spin polarization is described by $P(T) = P_0\left(1 - \alpha T^{3/2}\right)$. By substituting $P_1$ and $P_2$ in equation (2) with $P(T)$, the material-dependent constant $\alpha$ can be estimated to range from $1.1 \times 10^{-4}$ to $4.4 \times 10^{-4}$ $K^{-3/2}$. This value is slightly larger than that of $\alpha_{NiFe} = 3 \sim 5 \times 10^{-5}$ $K^{-3/2}$ found in ref. [40]. One possible reason for this difference may be the additional scattering to impurities as well as from the Au layer.

Further understanding on the physical mechanism leading to the MR is provided by *ab initio* density functional theory (DFT) based electron transport calculations for the Py/$MoS_2$/Py junction setup shown in Figure 4a (see Computational Methods for details). We find that the minimum energy occurs at a distance $d$ = 2.1 Å between the Py and $MoS_2$ layers, and this geometry was used for all the calculations. Transport results reveal that the single $MoS_2$ layer junction is metallic and presents a large conductance of the order of $1.2 \times 10^{14}$ S/m$^2$. The spin resolved transmission, $T$, as a function of the energy, $E$, with respect to the Fermi level, $E_F$, are shown in Figure 4b for both the parallel and antiparallel configurations of the junction. For the parallel alignment the transmission for the up spins, $T_\uparrow$, is larger than that for the down-spins, $T_\downarrow$, for energies around $E_F$. Combined with the fact that the density of states (DOS) in Py for up-spins is smaller than the one for down-spins (see supplementary Figures S4b and S4c), this shows that the Py/$MoS_2$ junction acts as spin-filter, where up-spin electrons get transmitted more likely than down-spin electrons. Interestingly $T_\uparrow$ is a smooth function of $E$ for $E - E_F > -1$ eV, since for those energies the states in the Py have *s* type character. Moreover $T_\uparrow$ increases for larger energies, following the rise in the Mo density of states (see supplementary Figure S4a). In contrast, $T_\downarrow$ is less smooth due to the fact that the down-spin *d* states of the Py extend across a wide energy range around $E_F$. The small peak $T_\downarrow$ around $E_F$ is associated to an interface state arising between the Py and the $MoS_2$ layers (see supplementary Figure S4c). The transmissions for up and down spins in the antiparallel configuration are nearly identical (a small asymmetry arises because our transport setup does not possess an exact inversion symmetric along *z*), and $T$ at $E_F$ is intermediate between $T_\uparrow$ and $T_\downarrow$ of the parallel alignment. Overall we find a low bias magnetoresistance, $MR = \frac{T_{parallel}(E) - T_{antiparallel}(E)}{T_{antiparallel}(E)} \approx 9\%$.

Here, we note that $MoS_2$ monolayers are known to have vacancies in the S layers[41,42], which might act as active scattering centers. As a consequence, the injected electrons can get scattered in the spacer layer and the spin polarization of the current is usually reduced. Inelastic scattering events from phonons or magnons will also generally reduce the polarization of the current. Therefore, while we expect the ballistic transport scenario for defect-free $MoS_2$ junctions to capture the main physics of the contact structures and their transport characteristics, it is expected to overestimate the experimental MR ratios, in an analogous way to what found for the extensively studied Fe/MgO junctions[12-16]. Moreover, the MR can also be potentially reduced due to the rather strong spin-orbit coupling of the monolayer $MoS_2$. However as shown in Ref. 19 the inclusion of spin-orbit effects in the calculations has only a marginal effect on the resulting MR. Taking into account these additional factors our simulated results are in qualitative agreement with the experimental findings, and can be considered as an upper bound to the possible magnetoresistance with permalloy electrodes. Note that in experiment the junctions are likely not single crystals as used in the simulations, but will have a polycrystalline character. Generally such deviations from perfect crystallinity will lead to a reduction of the MR values, which is indeed consistent with the fact that the experimental MR is somewhat lower than the theoretical one for the ideal interfaces. In addition, the MR is significantly smaller than what we previously calculated for the Fe/$MoS_2$/Fe junctions[19], which is entirely due to the low polarization induced by the permalloy electrodes when compared to Fe. This indicates that the MR ratio can be drastically improved through careful choice of the magnetic electrode.

The ($k_x - k_y$)-resolved transmission coefficients at $E = E_F$ are presented in Figure 4c and d. For the parallel setup, the transmission is large across a wide area of the $k_x - k_y$ plane for the up spin electrons, whereas it is reduced for the down spin case. Note in particular that at the center of the Brillouin zone ($k_x = k_y = 0$), there is a suppression of the transmission for the down spins, although there are actually more open channels available in the Py electrodes for spin down than for spin up. This is a consequence of symmetry filtering, where only particular states couple across the junction and can be transmitted[19]. The sharp features in the down-spin transmission around the Brillouin zone center are due to an enhanced transmission through the interface states. For the antiparallel case the overall transmissions approximately corresponds to an average of the up- and down-spin transmission of the parallel case. Our theoretical calculations elucidate the nature of the spin filtering effect in $MoS_2$, and agree well with the experimental results.

In conclusion, we have demonstrated a spin-valve effect in monolayers $MoS_2$ sandwiched between Permalloy electrodes. The effect persists up to 240 K with a moderate MR. We anticipate that a much larger MR could be realized through optimizing the interfaces and by incorporating multi-layer $MoS_2$ spacers. Alternatively one can use other TMDs materials, such as $WSe_2$, as spacer. We are confident that our results will pave the way for further TMDs-based tunable nanoscale spintronic devices.

## Methods

Monolayer $MoS_2$ is grown in a CVD tube furnace by using high-purity molybdenum and sulphur as the source materials, similar to a previous report[43]. The bottom electrodes were fabricated by *e*-beam lithography (EBL). The matrix of lines as shown in Figure 1d are designed to increase the successful rate in the following $MoS_2$ transfer process. Subsequently, Py/Co electrodes were deposited by magnetron sputtering without breaking the vacuum. After removing it from the chamber, the substrate was directly immersed into acetone for lift-off. Then the monolayer $MoS_2$ was quickly transferred onto the bottom electrodes. All these process had to be as quickly as possible to avoid oxidation of the Py layer. Finally, the top electrodes were fabricated by another run of EBL and metal deposition process. Prior to the measurements, the devices were annealed at 380 K for two hours in vacuum to remove interface molecules between $MoS_2$ and ferromagnetic metal electrodes.

**Transfer methods**

We spin-coated a very thick layer of PMMA onto Si/$SiO_2$ substrate, which has a large amount of monolayer $MoS_2$ on top. The process was repeated several times to make the PMMA layer thick enough so that the polymer could be picked up directly by tweezers. Then we separated the polymer from the substrate by etching in a 30% KOH solution to make a suspended PMMA layer with $MoS_2$ underneath, and washed it using deionized water. Before transferring the polymer onto the prepared bottom electrodes, we put a droplet of isopropyl alcohol (IPA) on the chip to make the PMMA lie on the substrate unfolded. After drying the chip by the hotplate, the PMMA was finally removed by acetone, resulting in a chip containing FM/$MoS_2$ device structure.

**Computational Methods**

First-principles transport calculations were performed using the SMEAGOL code[44–46], which interfaces the non-equilibrium Green's function formalism with density functional theory as numerically implemented in the SIESTA package[47]. Norm-conserving pseudopotentials are used to describe the core electrons, and a double-ζ polarized basis set is employed. The local density approximation (LDA) to the exchange-correlation functional is used. We construct a rectangular supercell in the plane, with a 14.37 Å × 8.30 Å cross section so that the lattice mismatch between Py (111 surface) and $MoS_2$ is minimized to 3.4%. A 2×4 ($k_x - k_y$)-point grid is used for the self-consistent calculation, while transmission and densities of states are then obtained by integrating over a denser 60×120 ($k_x - k_y$)-point mesh.

## Acknowledgements

This work was supported by the National Young 1000 Talent Plan, Pujiang Talent Plan in Shanghai, National Natural Science Foundation of China (61322407, 11474058). Part of the sample fabrication was performed at Fudan Nano-fabrication Laboratory. S.S. and A.N. thank CRANN and AMBER for financial support. I.R. acknowledges financial support from the EU project ACMOL (FP7-FET GA618082). We thank Trinity Centre for High Performance Computing (TCHPC) and Irish Centre for High-End Computing (ICHEC) for the computational resources provided.

## Author contributions

F.X. conceived the ideas and supervised the overall research. W.W. fabricated the devices and carried out the device characterizations. A.N., K.D., I.R. and S.S. carried out the electron transport calculations and analysed the results. L.T. contributed to the transfer method. Y.J. grew the monolayer $MoS_2$ and took the PL and Raman measurements. Y.L. and X.Y. contributed to the electrodes deposition and the sample measurements. Y.W. provided critical guidance during experiments. W.W. and A.N. wrote the paper with help from all other co-authors. F.X. and S.S. reviewed the manuscript.

## Additional information

The authors declare no competing financial interests.

Supporting Information Available: Additional data for control sample and other devices, discussions on the metallic behavior of $MoS_2$, and detailed simulation method. This material is available free of charge via the Internet at http://pubs.acs.org.

Reprints and permissions information is available online at

http://pubs.acs.org/page/copyright/index.html. Correspondence and requests for materials should be addressed to F.X.

**Figure 1**

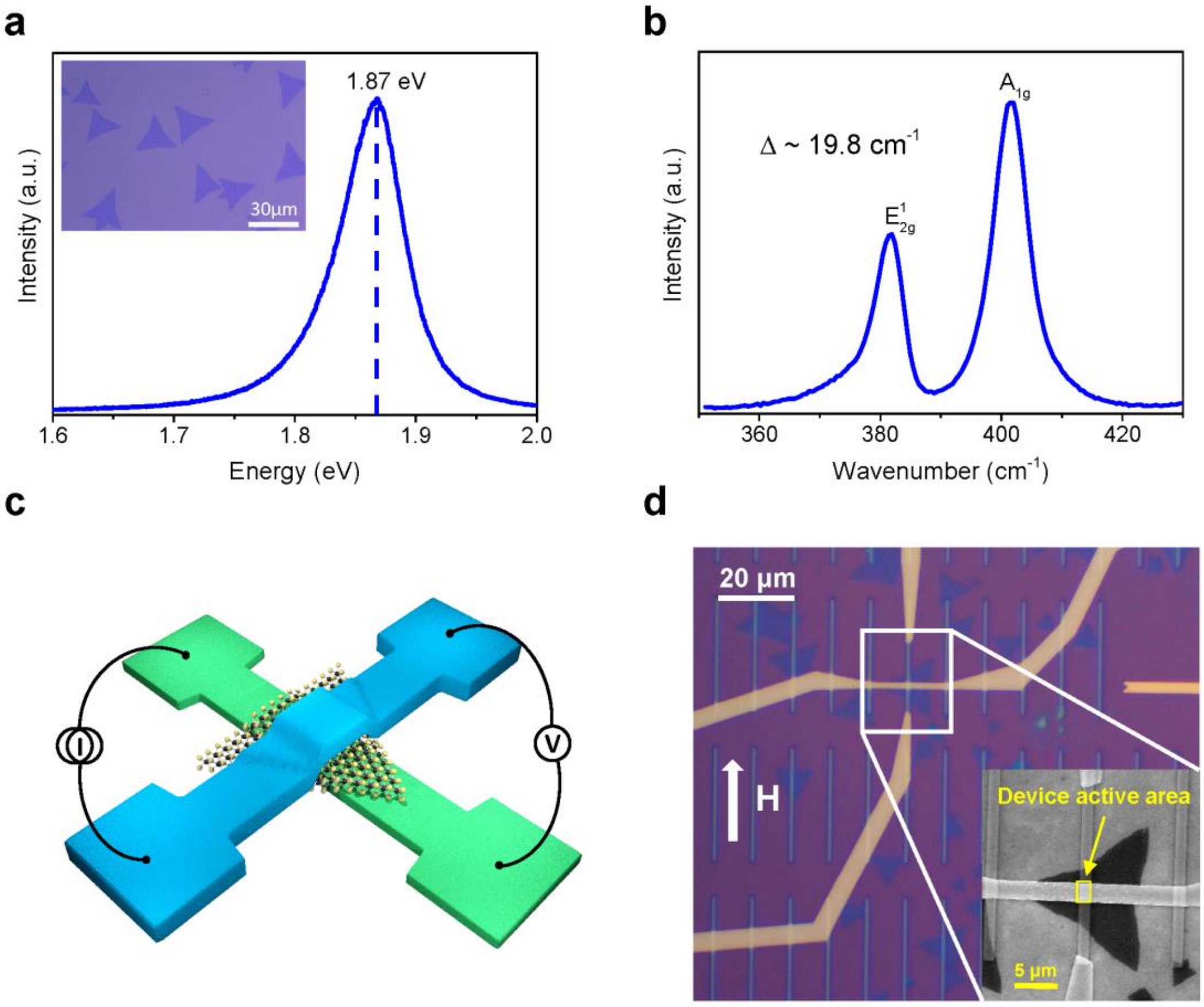

**Figure 1| Sample characterizations and device structure. a,** A typical photoluminescence (PL) spectrum of CVD-grown $MoS_2$. The peak position is determined to be 1.87 eV, indicative of monolayer $MoS_2$. The inset pictures the microscope image of the $MoS_2$ samples before the transfer process. **b,** A Raman spectrum of an as-grown monolayer $MoS_2$. The distance between two vibrating modes (in-plane mode $E^1_{2g}$ and out-of-plane mode $A_{1g}$) is ~ 19.8 $cm^{-1}$. **c,** A schematic diagram of the device. We apply 4-probe measurements where the current goes vertically through the junction area. As the electrodes are very thin in the range of 30~50 nm, the magnetization lies in-plane. **d,** A microscope and scanning electron microscope (SEM) (inset) image of the fabricated device. The magnetic field is applied in-plane.

**Figure 2**

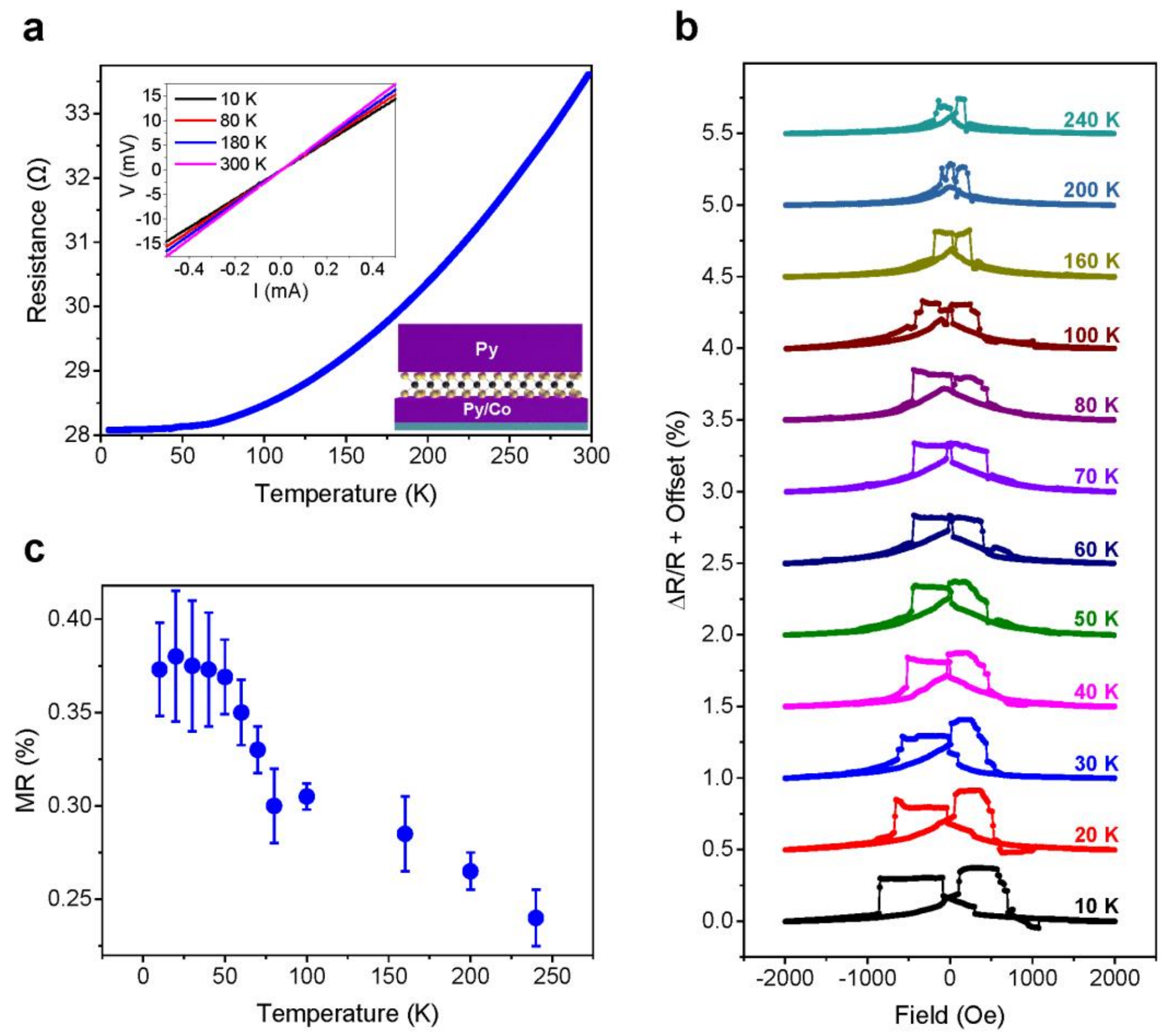

**Figure 2| Characterizations of the samples with Py/MoS2/Py/Co structure. a,** The junction resistance versus temperature. The resistance decreases with temperature, showing a metallic behavior. The inset is *I*-*V* curves for 4-probe measurement. The linear curve reveals an Ohmic contact between $MoS_2$ and electrodes. **b,** Magnetoresistance (MR) curves at different temperatures. The spin-valve effect is found up to 240 K, and the TMR ratio decreases as temperature rises up. **c,** MR magnitude as a function of temperature.

**Figure 3**

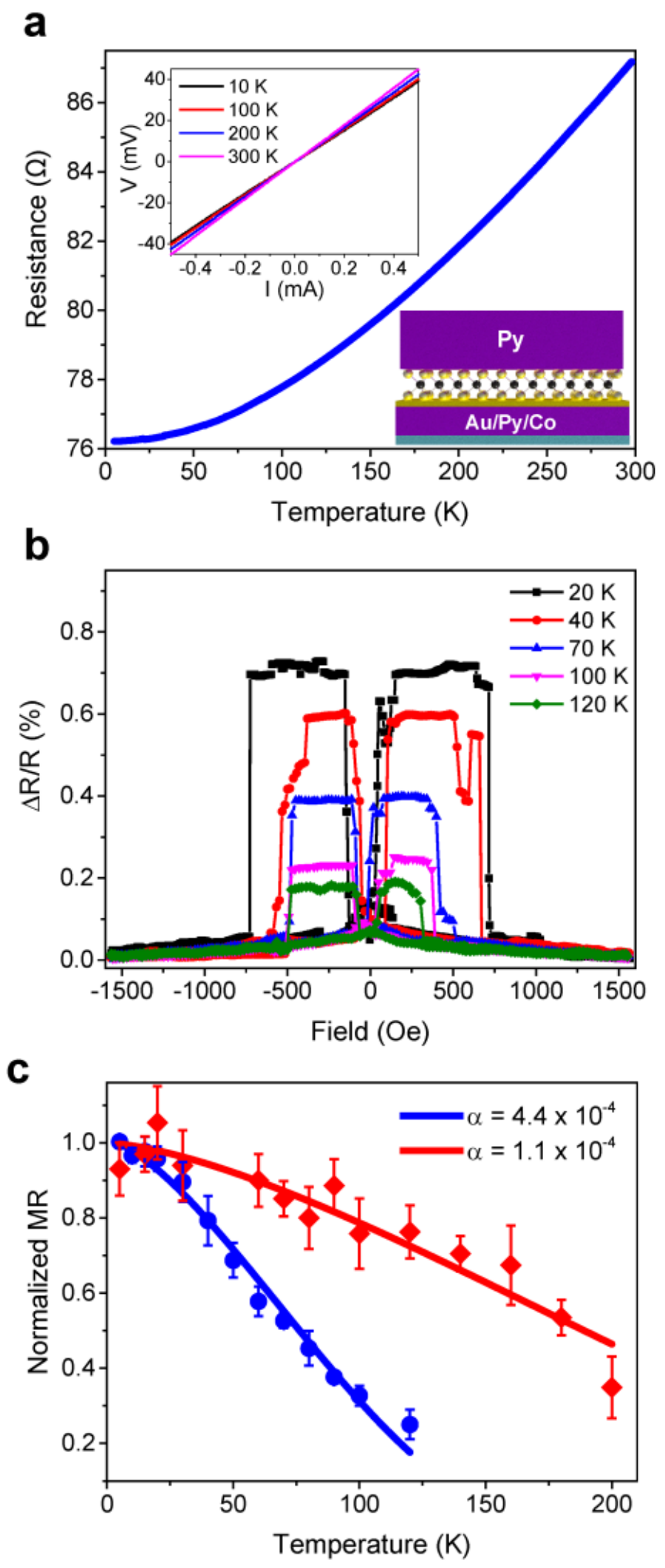

**Figure 3| Characterizations of the devices with Py/MoS2/Au/Py/Co structure. a,** The junction resistance versus temperature, showing a metallic behavior. The linear *I*-*V* curve (inset) exhibits an Ohmic contact between $MoS_2$ and electrodes. **b,** MR curves at different temperatures. The spin valve effect is found up to 120 K, and the MR ratio increases as the temperature is reduced. **c,** Normalized MR magnitude as a function of temperature for two junctions. The data is fitted to the Bloch's law, yielding an important material-dependent parameter $\alpha$ in the range of $1.1 \times 10^{-4}$ to $4.4 \times 10^{-4}$ $K^{-3/2}$.

**Figure 4**

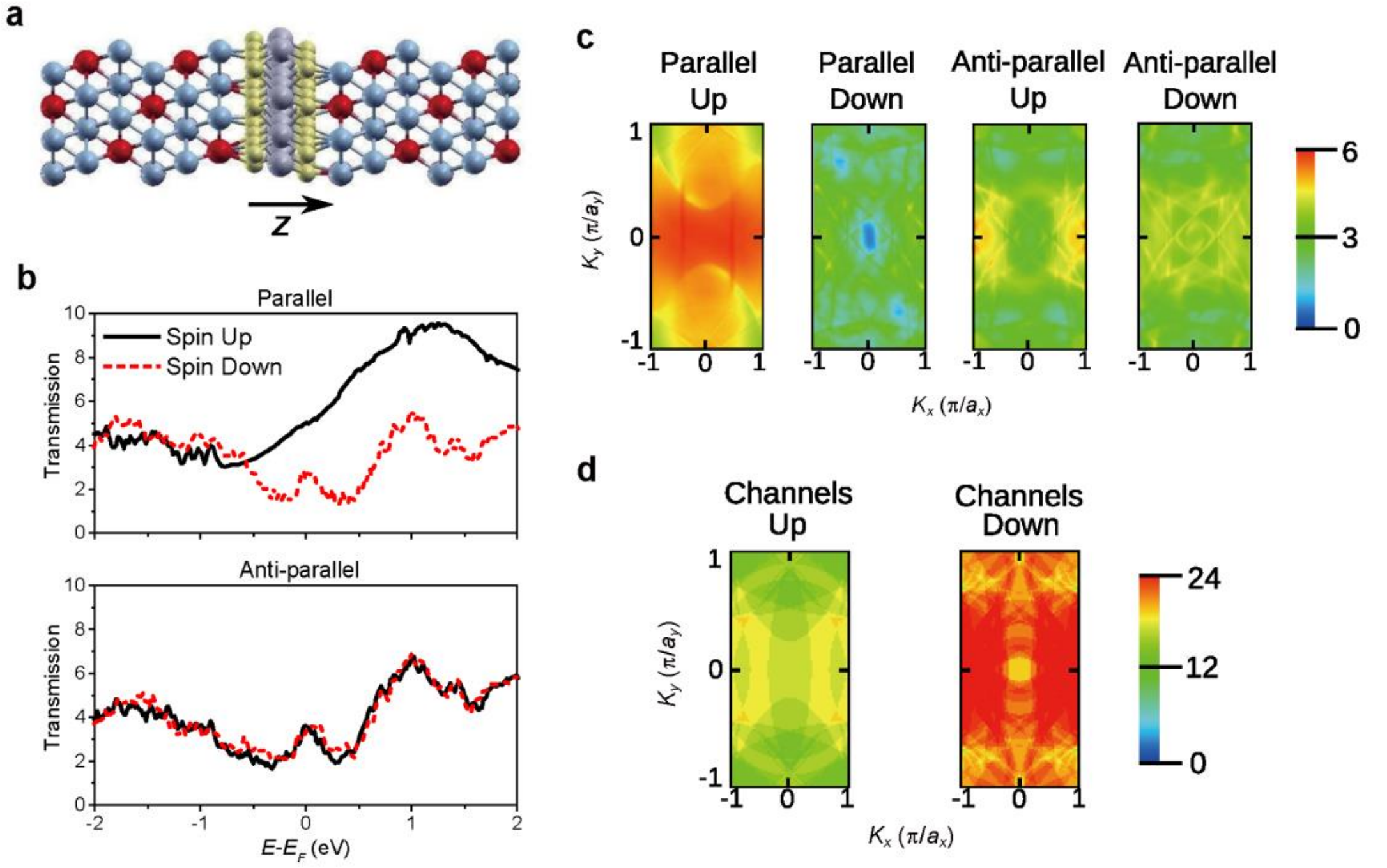

**Figure 4| Transport calculations for Py/MoS2/Py junction. a,** Scattering region for the first-principles calculations, with the transport along the *z* direction. Self-energies corresponding to semi-infinite Py electrodes are attached to the six left and right hand Py layers inside the scattering region. **b,** Transmission as a function of energy for spin up and spin down channels, for parallel and anti-parallel setup of the junction. **c,** $k_x - k_y$ resolved plot of the transmission at $E = E_F$ ($k_x$ running along the horizontal axis, and $k_y$ along the vertical axis), for the two spin directions for parallel and antiparallel configurations of the device. **d,** Spin up and spin down open channels for the electrodes at $E = E_F$, plotted across the $k_x - k_y$ plane.